\documentclass[%
 reprint,
 superscriptaddress,
 amsmath,amssymb,
 aps,
 prc,
 nofootinbib
]{revtex4-1}

\usepackage[utf8]{inputenc} 
\usepackage[colorlinks=true,allcolors=blue]{hyperref} 
\usepackage[toc,page]{appendix} 
\usepackage[shortlabels]{enumitem} 
\usepackage[dvipsnames]{xcolor}
\usepackage{soul}

\usepackage{graphicx} 
\usepackage{tabularx} 
\usepackage{dcolumn}  
\newcolumntype{d}[1]{D{.}{.}{#1}} 
\newcolumntype{R}{>{\raggedleft\arraybackslash}X} 

\usepackage{bm} 
\usepackage{mathtools} 
\usepackage{physics} 
\usepackage{simplewick} 

\begin{document}


\title{Global Description of Beta Decay with the Axially-Deformed Skyrme
Finite Amplitude Method: Extension to Odd-Mass and Odd-Odd Nuclei}

\author{E. M. Ney}
\email[]{evan.ney@unc.edu}
\affiliation{Department of Physics and Astronomy, CB 3255, University of North
Carolina, Chapel Hill, North Carolina 27599-3255, USA\looseness=-1}

\author{J. Engel}
\email[]{engelj@unc.edu}
\affiliation{Department of Physics and Astronomy, CB 3255, University of North
Carolina, Chapel Hill, North Carolina 27599-3255, USA\looseness=-1}

\author{N. Schunck}
\email[]{schunck1@llnl.gov}
\affiliation{Nuclear and Chemical Science Division, LLNL, Livermore, California
94551, USA\looseness=-1}

\date{\today}

\begin{abstract}
We use the finite amplitude method (FAM), an efficient implementation of the
quasiparticle random phase approximation, to compute beta-decay rates with
Skyrme energy-density functionals for 3983 nuclei, essentially all the
medium-mass and heavy isotopes on the neutron rich side of stability.  We employ
an extension of the FAM that treats odd-mass and odd-odd nuclear ground states
in the equal filling approximation.  Our rates are in reasonable agreement both
with experimental data where available and with rates from other global
calculations.  
\end{abstract}

\maketitle

\section{\label{sec:intro}Introduction}

The origin of elements heavier than iron still remains an open question. Early
work has shown that neutron capture in astrophysical processes is responsible
for synthesizing those elements~\cite{Burbidge1957,Meyer1994}. Rapid neutron
capture, through the ``$r$-process,'' is particularly interesting because its
astrophysical site is still uncertain.  The multi-messenger neutron star merger
GW170817~\cite{Abbott2017} recently provided evidence that such events are the
dominant source of $r$-process elements, but quantitative conclusions require
more data.  We need more reliable astrophysical simulations to connect future
multi-messenger events with details of the underlying nucleosynthesis.  

Abundances of $r$-process elements depend on a variety of nuclear properties,
including masses, neutron-capture cross sections, photo-disintegration cross
sections, fission yields, and beta-decay half-lives~\cite{Horowitz2019}.
Although some of these properties have been measured and
tabulated~\cite{ENSDF2019}, the majority of nuclei relevant for the $r$-process
are too unstable to be produced in the lab. Reliable $r$-process simulations
thus require calculations in neutron-rich nuclei.  Beta-decay half-lives are
particularly important because they determine the overall timescale for neutron
capture in the $r$-process~\cite{Moller1997,Engel1999} and affect the shape of
the final abundance pattern~\cite{Mumpower2014,Shafer2016}. 

A variety of global beta-decay calculations exist, in the semi-gross
theory~\cite{Nakata1997}, in a quasiparticle random-phase approximation (QRPA)
plus macroscopic finite-range droplet model (FRDM)
approach~\cite{Moller1997,Moller2003}, in covariant density functional theory
(DFT)~\cite{Marketin2016}, etc.  DFT, covariant or not, is particularly
attractive because it offers a self-consistent, microscopic framework for
computing properties across the nuclear chart~\cite{Ring2004,Schunck2019}. For
the calculation of beta decay in deformed superfluid nuclei, DFT amounts to the
QRPA, built on a ground state produced by the Hartree-Fock-Bogoliubov (HFB)
method, which incorporates pairing correlations into mean fields, all with
density-dependent interactions.

In odd-mass and odd-odd nuclei (hereafter ``odd'' nuclei) pairing is ``blocked''
and the HFB ground-state contains a quasiparticle excitation~\cite{Ring2004}.
This complicates calculations because the ground-state is no longer invariant
under time reversal~\cite{bertsch2009a,Schunck2010}. As a result additional
approximations are often made in beta-decay calculations.
Reference~\cite{Homma1996}, for example, treats one-quasiparticle states
perturbatively, while Ref.~\cite{Marketin2016} treats them as if they were
zero-quasiparticle states. A more consistent way to approximate HFB blocked
states while preserving time-reversal symmetry is through the equal filling
approximation (EFA)~\cite{Perez-Martin2008}. Numerous studies showed that the
EFA is an excellent approximation to exact
blocking~\cite{duguet2001a,bertsch2009,Schunck2010}.
Reference~\cite{Shafer2016} recently developed a method to extend the EFA to the
QRPA. 

In this work we use the extension to carry out a global calculation of
allowed and first-forbidden contributions to beta-minus decay in odd nuclei from
near the valley of stability out to the neutron drip line. We use a global
Skyrme density functional determined in Ref.~\cite{Mustonen2016}, thus extending
that work, which was restricted to even-even nuclei, to all isotopes that play a
role in the $r$-process.

This rest of this paper is as follows: Sec.~\ref{sec:theory} presents background
for the finite amplitude method (FAM), which we use to compute QRPA strength
functions, and its extension to the EFA.  Section~\ref{sec:method} outlines some
improvements to our implementation of the FAM since the work of
Ref.~\cite{Mustonen2016}. Section~\ref{sec:results} presents our results, compares
them to those of other papers and to experiment, and addresses subtleties of
the EFA-FAM.  Section~\ref{sec:conclusions} contains concluding remarks.  

\section{\label{sec:theory}The proton-neutron finite amplitude method (pnFAM)}
\subsection{\label{ssec:pnfam}The pnFAM for pure states}

The QRPA linear-response function is the same as that from time-dependent HFB
theory~\cite{Ring2004}. One way of computing it is to diagonalize a set of
matrices with dimension equal to that of the two-quasiparticle space.  The
construction of these matrices, which require two-body matrix elements of the
potential, is time consuming in deformed nuclei.  The FAM sidesteps the
matrices, significantly speeding up the computation of linear response produced
by energy-density functionals.  Reference~\cite{Nakatsukasa2007} first presented
the FAM for the ordinary RPA, and Ref.~\cite{Avogadro2011} did the same for the
QRPA.  Since then, the method has been used with covariant density
functionals~\cite{Niksic2013,Liang2013,Liang2014} and employed to compute
transition strength in several
contexts~\cite{Inakura2009a,Inakura2009,Inakura2010,Stoitsov2011,Oishi2016}.

Here we build on the work of Refs.~\cite{Mustonen2014,Mustonen2016,Shafer2016},
which used a charge-changing version of the FAM called the pnFAM together with
the contour-integral method of
Refs.~\cite{Nakatsukasa2014,Hinohara2015,Hinohara2013,Hinohara2015a} to compute
beta-decay rates.  A detailed account of the pnFAM and its application to beta
decay appears in Ref.~\cite{Mustonen2014}.  Reference~\cite{Shafer2016} used
the EFA to extend the pnFAM to odd nuclei
and compute beta-decay rates in the rare-earth nuclei that are important for
$r$-process simulations.  In order to highlight a few subtleties of the
EFA-pnFAM, we recapitulate the main points of the theory here.

We begin with the time-dependent HFB equations
\begin{equation}
\label{eq:tdhfb}
i \dot{\mathbb{R}}(t) = \big[\mathbb{H}[\mathbb{R}(t)]+\mathbb{F}(t),
\ \mathbb{R}(t) \big]\,.
\end{equation}
Here, $\mathbb{R}$ is the generalized HFB density matrix, $\mathbb{H}$ is the
HFB Hamiltonian matrix, and $\mathbb{F}$ is a matrix that represents a one-body
time-dependent perturbation. The blackboard-bold letters indicate that these
matrices are in the HFB quasiparticle basis, defined by the Bogoliubov
transformation $\mathbb{W}$:
\begin{equation}
\label{eq:hfb_W}
\mathbb{W} = 
\begin{pmatrix}
 U & V^* \\ 
 V & U^*  
 \end{pmatrix}
\,, 
 \end{equation}
where $U$ and $V$ are themselves matrices.  In this basis the static
ground-state Hamiltonian and the associated generalized density are diagonal:
\begin{equation}
\label{eq:hfb_ground_state}
\mathbb{H}_{0}  = 
 \begin{pmatrix}
 E & 0 \\ 0 & -E
 \end{pmatrix}
 \,, \quad \mathbb{R}_0 = 
 \begin{pmatrix}
 0 & 0 \\ 0 & 1
 \end{pmatrix}
 \,. 
\end{equation}

To first order in the perturbation $\mathbb{F}$, Eq.~\eqref{eq:tdhfb} is 
\begin{equation}
\label{eq:linear_response}
 i \dot{\delta \mathbb{R}}(t) = \big[\mathbb{H}_0, \delta \mathbb{R}(t) \big] +
 \big[\delta\mathbb{H}(t) + \mathbb{F}(t), \mathbb{R}_0 \big] \,, 
\end{equation}
with $\delta \mathbb{R}(t) = \mathbb{R}(t) - \mathbb{R}_0$.  If the perturbation
is harmonic, the time-dependent quantities $\mathbb{F}(t)$, $\delta
\mathbb{H}(t)$, and $\delta \mathbb{R}(t)$ all take the form (e.g.\ for
$\mathbb{F}$)
\begin{equation}
\label{eq:time_dependent_quantities}
\begin{aligned}
 \mathbb{F}(t) &= \mathbb{F}(\omega) e^{-i\omega t} + \mathbb{F}^\dagger(\omega)
 e^{i\omega t} \\
 \mathbb{F}(\omega) &=
 \begin{pmatrix}
 {F}^{11}(\omega) & {F}^{02}(\omega) \\ 
 -{F}^{20}(\omega) & -{F}^{\overline{11}}(\omega)
 \end{pmatrix} 
 \,.
\end{aligned}
\end{equation}
We denote the perturbed density more specifically by 
\begin{equation}
\label{eq:density_response_omega}
 \delta \mathbb{R}(\omega) =
 \begin{pmatrix}
 P(\omega) & X(\omega) \\
 -Y(\omega) & -Q(\omega)
 \end{pmatrix} \,.
\end{equation}
When one substitutes Eqs.~\eqref{eq:time_dependent_quantities}
and~\eqref{eq:density_response_omega} into Eq.~\eqref{eq:linear_response}, the
diagonal blocks $P$ and $Q$ vanish, and for a charge-changing external field
only the proton-neutron matrix elements of the response are nonzero. These
conditions lead to the pnFAM equations
\begin{equation}
\label{eq:pnfam} \begin{aligned}
 &\big( E_\pi + E_\nu - \omega \big) X_{\pi\nu}(\omega) = - \big( \delta
 {H}^{20}_{\pi\nu}(\omega) + {F}^{20}_{\pi\nu}(\omega) \big) \\
 &\big( E_\pi + E_\nu + \omega \big)\ Y_{\pi\nu}(\omega) = - \big( \delta
 {H}^{02}_{\pi\nu}(\omega) + {F}^{02}_{\pi\nu}(\omega) \big) \,, 
\end{aligned}
\end{equation}
where the label $\pi$ denotes protons and the label $\nu$ denotes neutrons.  The
use of a finite-difference method to compute $\delta {H}$ is the source of the
FAM's speed.  Because we do not consider mixing of protons and neutrons in the
underlying HFB ground state, and because Skyrme functionals in use depend at
most quadratically on charge-changing densities, the finite difference in the
pnFAM reduces exactly to the evaluation of the Hamiltonian with the perturbed
densities:
\begin{equation}\label{eq:hamiltonian_response}
\begin{aligned}
    \delta \mathbb{H}^{(pn)} &= \lim_{\eta \to 0} \frac{1}{\eta} \bigg(
    \mathbb{H} \Big[\mathbb{R}_0^{(pp,nn)} + \eta \delta\mathbb{R}^{(pn)} \Big]
    - \mathbb{H} \Big[\mathbb{R}_0^{(pp,nn)} \Big] \bigg) \\
    &= \mathbb{H} \Big[\delta\mathbb{R}^{(pn)} \Big] \,.
\end{aligned}
\end{equation}

Once the FAM amplitudes $X$ and $Y$ are known, one can compute the strength
function:
\begin{equation}
\label{eq:fam_strength_1}
\frac{dB(F,\omega)}{d\omega} = -\frac{1}{\pi} \Im S(F, \omega) \,, 
\end{equation}
where
\begin{equation}
\label{eq:fam_strength_2}
\begin{aligned}
 S(F, \omega) &= \sum\limits_{\pi\nu} \big[ {F}^{20^*}_{\pi\nu}
 X_{\pi\nu}(\omega) + {F}^{02^*}_{\pi\nu} Y_{\pi\nu}(\omega) \big]\\
 &= - \sum\limits_n \bigg( \frac{\lvert \bra{n} \hat{F} \ket{0} \rvert^2}{\Omega_n -
 \omega} + \frac{\lvert \bra{n} \hat{F}^\dagger \ket{0} \rvert^2}{\Omega_n + \omega}
 \bigg) \,. 
\end{aligned}
\end{equation}

The FAM strength function has poles at QRPA excitation energies $\Omega_n$ with
residues equal to the transition probabilities $\lvert \bra{n} \hat{F} \ket{0}
\rvert^2$.  It also contains poles at $-\Omega_n$, with residues equal to the
negative of transition probabilities for the conjugate operator $\lvert \bra{n}
\hat{F}^\dagger \ket{0} \rvert^2$. In beta-minus-decay calculations $\hat{F}$
contains the isospin lowering operator and $\hat{F}^\dagger$ contains the
isospin raising operator; cf. Ref.~\cite{Mustonen2014} for a list of the six
allowed and first-forbidden operators. Thus, the poles with positive and
negative residues correspond to beta-minus and beta-plus transitions,
respectively. This point will become important in the EFA-pnFAM.

In practice we construct the strength function by solving the pnFAM equations
separately for each of a large set of complex frequencies $\omega$. From
Eqs.~\eqref{eq:fam_strength_1} and~\eqref{eq:fam_strength_2}, it is
straightforward to show that each pole of $S(F, \omega)$ on the real axis
contributes a Lorentzian of half-width $\gamma = \Im[\omega]$ to the strength
function in the complex plane. The strength may be be calculated for a set of
frequencies close to the real axis with a fixed half-width to mimic experimental
strength measurements, or along a closed contour in the complex plane to
calculate cumulative strength or decay rates.

\subsection{\label{ssec:ft_pnfam}The pnFAM for statistical ensembles}

Many HFB codes use the EFA to avoid the difficulties associated with the
breaking of time-reversal symmetry~\cite{Ring2004,bertsch2009a} in odd nuclei.
The originally \emph{ad hoc} EFA can be understood as a special case of
statistical HFB theory for an ensemble that is symmetric under time
reversal~\cite{Perez-Martin2008,Schunck2010}. In systems with time-reversal
symmetry, a state $\ket{\lambda}$ and its time-reversed partner
$\ket{\overline{\lambda}}$ are degenerate, and the equal filling quasiparticle
occupation probabilities, for axial but not spherical symmetry, are
\begin{equation}
\label{eq:efa_occupations}
f_{\mu\nu} = \frac{1}{2} ( \delta_{\nu\lambda} + \delta_{\nu\overline{\lambda}})
\delta_{\mu\nu} \,. 
\end{equation}
In odd-odd nuclei, both the odd-proton and odd-neutron quasiparticles have
non-zero occupation probabilities. Note that in this work, we do not consider
neutron-proton pairing at the HFB level.

The statistical extension of the QRPA~\cite{Sommermann1983} lets us use the FAM
to treat excitations of HFB ensembles, taking into account at least partially
the polarization of the even-even ``core'' by the odd nucleon.  The EFA-FAM can
be derived in the same way as the ordinary FAM, by promoting the ground-state
generalized density matrix to a statistical density operator.  Expectation
values that, for example, define the particle densities, then become ensemble
averages. The generalized HFB density matrix is no longer a projector and takes
the more general form
\begin{equation}
\label{eq:ft_density}
\widetilde{\mathbb{R}}_0 =
\begin{pmatrix}
f & 0 \\ 0 & 1-f
\end{pmatrix} \,. 
\end{equation} 
In the usual finite-temperature theory, based on the grand canonical ensemble,
the occupation probabilities are given by $f_{\mu\nu}= {(1+\exp(\beta
E_\mu))^{-1}} \delta_{\mu\nu}$~\cite{Goodman1981}. In the EFA we impose the
occupation probabilities of Eq.~\eqref{eq:efa_occupations}.

To obtain the statistical pnFAM equations we simply replace the ground-state
generalized density of Sec.~\ref{ssec:pnfam} with that of Eq.\
\eqref{eq:ft_density}. The diagonal elements of the density response no longer
vanish, and new statistical factors appear.  Once again, for a charge-changing
perturbation we need only the proton-neutron matrix elements, and so the
statistical pnFAM equations are
\begin{equation}
\begin{aligned}
&\big( E_\pi - E_\nu - \omega \big) P_{\pi\nu}(\omega) = - (f_\nu - f_\pi) \big(
\delta {H} + {F} \big)^{11}_{\pi\nu}(\omega) \\
&\big( E_\pi + E_\nu - \omega \big) X_{\pi\nu}(\omega) = - (1 - f_\pi - f_\nu)
\big( \delta {H} + {F} \big)^{20}_{\pi\nu}(\omega) \\
&\big( E_\pi + E_\nu + \omega \big) Y_{\pi\nu}(\omega) = - (1 - f_\pi - f_\nu)
\big( \delta {H} + {F} \big)^{02}_{\pi\nu}(\omega) \\
&\big( E_\pi - E_\nu + \omega \big) Q_{\pi\nu}(\omega) = - (f_\nu - f_\pi) \big(
\delta {H} + {F} \big)^{\overline{11}}_{\pi\nu}(\omega) \,. 
\end{aligned}
\end{equation}
The additional $P$ and $Q$ amplitudes arise because the non-zero occupation
probabilities allow quasiparticles to be destroyed as well as created.  The new
transitions introduce an additional set of QRPA eigenvalues that contain
quasiparticle-energy differences rather than sums~\cite{Sommermann1983}. It is
possible for these energy differences to be negative, indicating a transition to
a state of lower energy.  This does not mean, however, that the QRPA fails, as
it does when the eigenvalues are imaginary.  The statistical FAM strength has
the same form as the usual strength in Eq.~\eqref{eq:fam_strength_2}, but the
residues become ensemble-averaged transition strengths, and $n$ runs over the
expanded set of QRPA modes.  More details on the EFA-FAM and a demonstration
that it includes all necessary transitions for odd states, in the context of the
particle-rotor model~\cite{Bohr1998}, appear in Ref.~\cite{Shafer2016}.

\section{\label{sec:method}Computational method}
\subsection{\label{ssec:groundstates}HFB ground states and functional}

In obtaining our global set of half-lives, we introduce a number of small
improvements to the procedure of Ref.~\cite{Mustonen2016}, in addition to the
changes required to compute half-lives of odd nuclei. The first is in the
determination of the HFB ground state/ensemble. To make sure that we identify 
the correct ground state, we perform three different calculations for each 
even-even nucleus by constraining the first ten iterations of the HFB solver to 
an oblate, spherical and prolate quadrupole shape before releasing the 
constraint. In contrast to Ref.~\cite{Mustonen2016}, which used a set of three 
fixed quadrupole constraints for all nuclei, we use the first-order 
mass-dependent relation~\cite{Ring2004}
\begin{equation}\label{eq:Q2}
Q_2 = \frac{5}{100 \pi} \beta_2 A^{5/3} \,, 
\end{equation}
with values $\beta_2 = -0.2, 0.0, +0.2$. This procedure gives one, two or three 
different deformed minima, depending on the even-even nucleus. We then identify 
a number of candidate quasiparticle states within 1~MeV of the Fermi surface to 
block in the EFA. For odd-odd nuclei we consider all possible combinations of 
proton and neutron candidates. For every candidate (or candidate pair), we 
carry out the EFA on top of each available deformed even-even core, without 
constraints, and select the solution with the lowest energy.  On occasion these 
are meta-stable super-deformed states, which we discard.

We use the Skyrme functional SKO$'$~\cite{Reinhard1999}, which was found in
Ref~\cite{Mustonen2016} to give accurate $Q$-values across the nuclear chart.
We fit the like-particle pairing strengths to the experimental pairing gaps of
ten isotopes picked in a wide mass range $50 \le A \le 230$, and apply an
ulta-violet cutoff of 60~MeV to the single particle space. For the pnFAM portion
of the calculation we set the time-odd parameters and isoscalar pairing strength
to the values determined in the fit ``1A'' of that reference.  We therefore also
use the same 16-shell deformed harmonic-oscillator basis that was used in the
original fit. All HFB calculations are performed with the latest version of the
\textsc{hfbtho} code~\cite{perez2017}.

\subsection{\label{ssec:betadecay}Beta-decay half-lives}

The next set of changes concerns the computation of the beta-decay half-lives,
which is discussed in detail in Ref.~\cite{Mustonen2014}. The procedure therein
allows us to sum the phase-space-weighted strengths to all energetically allowed
daughter states. For allowed transitions, we obtain the rate and half-life via
\begin{equation}\label{eq:halflife}
    \lambda = \frac{\ln2}{\kappa} \sum\limits_n f(W_n) \lvert \bra{n} F \ket{0}
    \rvert^2, \quad t_{1/2} = \frac{\ln2}{\lambda} \,,
\end{equation}
where $\ket{n}$ is the $n^{\text{th}}$ state in the daughter nucleus,
${W_n=E_n/m_e c^2}$ is the energy, in units of electron mass, of the electron
emitted during a transition to that state, and $\kappa = 6147.0 \pm 2.4 s$.  To
include first-forbidden transitions, we must consider a more complicated
phase-space-weighted ``shape factor''~\cite{Mustonen2014}.  We evaluate the
right side of the first relation in Eq.~\eqref{eq:halflife} by integrating the
phase-space-weighted strength (Eq.~\eqref{eq:fam_strength_2}) along a circular
complex energy contour~\cite{Mustonen2014} that encloses all the poles below the
decay $Q$-value.  Because the phase-space integral $f(W_n)$ is not analytic, the
authors of Ref.~\cite{Mustonen2014} fit a polynomial to the integrals on the
real axis, and analytically continued the polynomial.  High-degree polynomials
on evenly spaced grids, however, exhibit the Runge-phenomenon~\cite{Runge1901},
and can oscillate rapidly in the complex plane.  We therefore elect here to use
a rational function to interpolate the phase-space integrals on a 20-point
Chebychev grid.  Because the contour integrand is quite smooth, we use
Gauss-Legendre quadrature to perform the contour integration.

The maximum QRPA energy relevant for beta decay defines the right bound of the
circular energy contour.  With the treatment of $Q$-values in
Refs.~\cite{Engel1999,Mustonen2014,Shafer2016}, the energy released in the
transition to the $n^{th}$ excited state in the daughter nucleus is
\begin{equation}
\label{eq:Qval}
Q_{\beta}^{(n)} = \Delta M_{n-H} + \lambda_n - \lambda_p - \Omega_{n} \,,
\end{equation} 
where $\Delta M_{n-H}$ is the neutron-hydrogen mass difference, $\lambda_{p}$
and $\lambda_{n}$ are the proton and neutron HFB Fermi energies, and $\Omega_{n}
$ ($n\geq 1$) is the excitation energy of the $n^{\rm th}$ QRPA mode above the
initial-nucleus ground state, after adjustment by the Fermi energies for the
change in particle number. (Note that $\Omega_1$ is the ``excitation energy'' of
the ground state of the daughter nucleus.) The maximum QRPA energy, which
corresponds to an energy release of zero is then the excitation energy of, e.g.,
the daughter ground state plus the energy released in the transition to that
state, 
\begin{equation}
E_{\rm max}^{\rm QRPA} 
= Q_{\beta}^{(1)} + \Omega_{1}
= \Delta M_{n-H} + \lambda_n - \lambda_p \,.
\label{eq:eqrpamax} 
\end{equation}
and can be evaluated without knowing the daughter ground-state energy itself.

The left bound of the circular energy contour must still be chosen. It must be
less than $\Omega_1$, which we do not know exactly, to include all relevant
poles in the response.  For even-even parent nuclei we can always choose it to
be zero because pairing correlations always make $\Omega_1$ positive.  For odd
parent nuclei, however, $\Omega_1$ can be negative.  If we neglect the effects
of the QRPA residual interaction, we find explicitly that 
\begin{equation}
\label{eq:Egs} 
\begin{aligned} 
\Omega_1^{\text{even}}    &\approx E_{\pi}^{\text{smallest}} + E_{\nu}^{\text{smallest}}\\
\Omega_1^{\text{n-odd}}   &\approx E_{\pi}^{\text{smallest}} - E_{\nu}^{\text{blocked}}\\ 
\Omega_1^{\text{p-odd}}   &\approx E_{\nu}^{\text{smallest}} - E_{\pi}^{\text{blocked}}\\
\Omega_1^{\text{odd-odd}} &\approx \text{min}\big[\Omega_1^{\text{p-odd}},\ \Omega_1^{\text{n-odd}} \big] \,.
\end{aligned} 
\end{equation}
The fact that $\Omega_1$ can be negative makes it difficult to choose the left
bound. If we expand the contour arbitrarily, we risk including beta-plus poles
with non-negligible negative strength\footnote{Poles are symmetric around zero,
so as soon as beta-minus strength appears at negative energy, some beta-plus
strength (inverted in sign) appears at positive energy.}, but if we do not
expand it enough, the QRPA residual interaction places $\Omega_1$ outside the
contour.  Because the pnFAM produces the strength function in
Eq.~\eqref{eq:fam_strength_1} directly, we do not have access to the underlying
QRPA eigenvectors and therefore cannot separate beta-minus poles from beta-plus
poles.  Both the inclusion of beta-plus poles or the accidental exclusion of
beta-minus poles at negative energies can cause the contour integration to
artificially reduce the integrated (and phase-space-weighted) beta-minus
strength, and therefore artificially increase the half-lives.  For lack of a
better prescription, we initially choose the left bound of the contour to be
\begin{equation}
\label{eq:contour_left_bound} 
E^{\text{QRPA}}_{\text{min}} =
\text{min}\big[0, \Omega_1] \,,
\end{equation}
with $\Omega_1$ given by the approximations in Eq.\ \eqref{eq:Egs}, but correct
the rates as described below when the contour integration appears to lead to
errors.

\begin{figure*}[t]
\includegraphics[width=2\columnwidth]{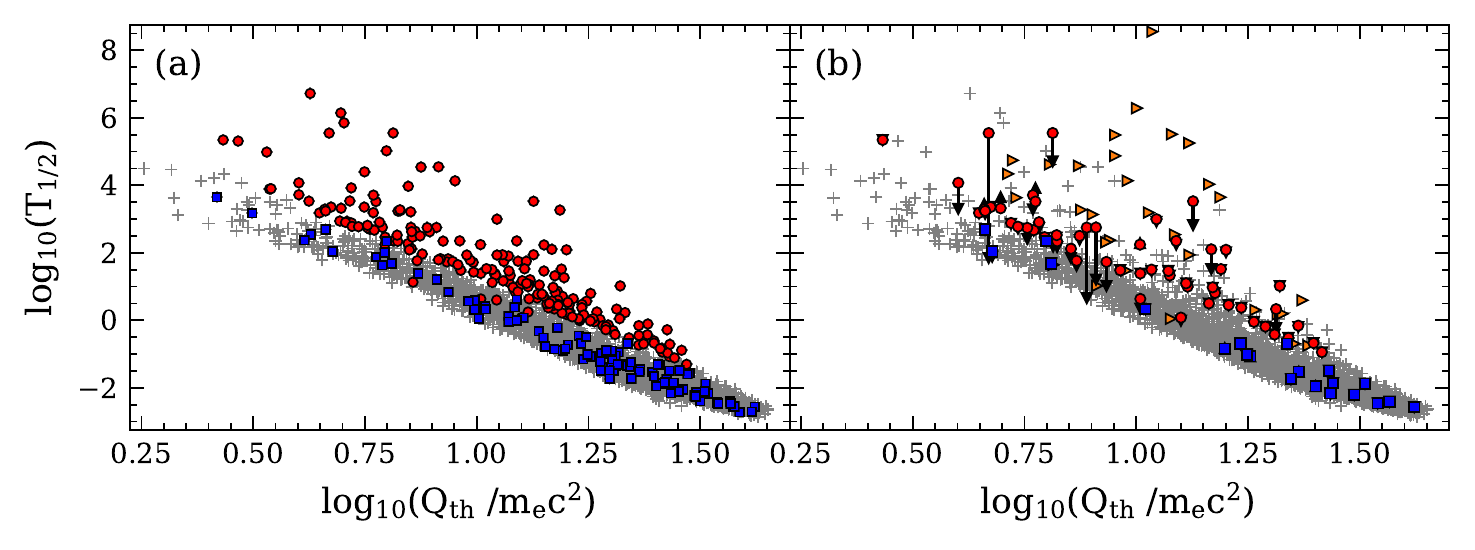}
\caption{\label{fig:hl_corrections} Corrections of half-lives in odd nuclei.
Half-lives for which we calculate strength functions are in panel a). Red
circles are for the 224 suspicious nuclei and blue squares for the 100 nuclei
from the random sample. Panel b) shows half-lives that are corrected as
described in the main text, with corrected values indicated by black arrowheads.
Orange triangles are corrected versions of originally negative half-lives.}
\end{figure*}

\begin{figure*}
\includegraphics[width=2\columnwidth]{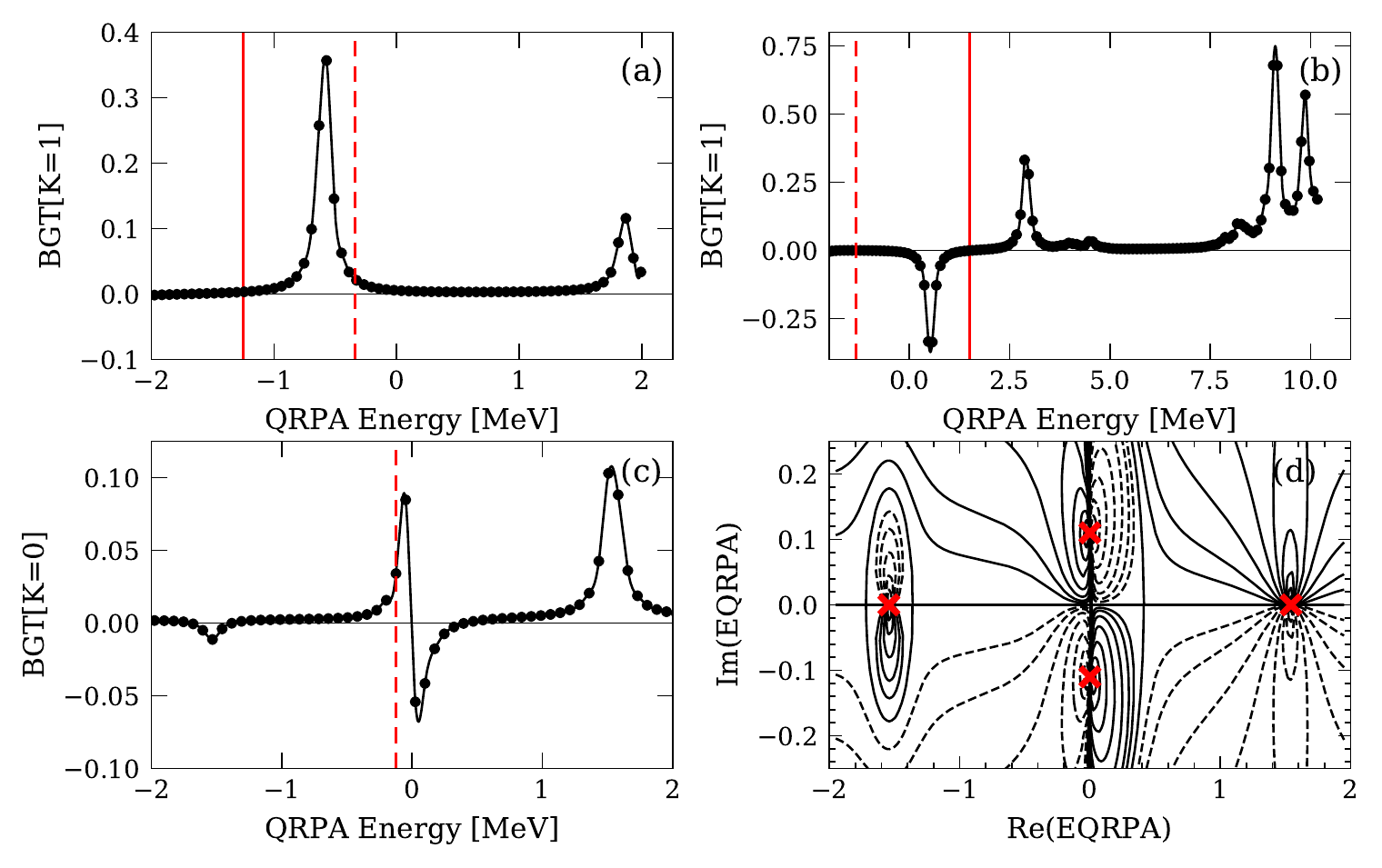}
\caption{\label{fig:contour_integration_failure} Examples in which contour
integration fails. The dashed vertical line indicates the left bound of the
contour from Eq.~\eqref{eq:Egs}, while the solid line indicates the adjusted
bound required to correct the rate.  Panels a) and b) show the Gamow-Teller
(K=1) strength function for $^{63}$Co and $^{53}$Sc, respectively.  Panel c)
illustrates an imaginary pole in the response function of $^{173}$Er, and d)
shows the response in the full complex plane, indicating poles near $\omega= \pm
0.1 i$.}
\end{figure*}

\section{\label{sec:results}Results} \subsection{\label{ssec:odd_corrections}
Half-lives and odd-nucleus subtleties}

To carry out our calculations we bundle the HFB code \textsc{hfbtho} and the
charge-changing FAM code \textsc{pnfam} together with a controlling python code
called \textsc{p$_{\textsc{Y}}$nfam}.  We calculate the beta-minus decay
half-lives of nuclei on the neutron rich side of stability, from $Z=20$ to
$Z=110$, out to the one-neutron drip line. The lightest nuclei in each isotopic
chain are near $A=50$, and coincide with those used in the global even-even
calculation of Ref.~\cite{Mustonen2016}.  We obtain 3983 ground states, 2998 of
which are odd isotopes.  Reference~\cite{Mustonen2016}, which included results
to the two-neutron drip line, obtained 1387 even-even ground states with the
same functional, versus our 985.  Our computation consumed roughly 270,000 Xeon
core hours. 

Our results in even-even nuclei agree very closely with those of
Ref.~\cite{Mustonen2016}, with a few improvements that can be attributed to our
updated procedures.  As mentioned in Sec.~\ref{ssec:betadecay}, however, our
contour-integration result may be inaccurate in odd nuclei if $\Omega_1$ is less
than zero. To assess the validity of the contour integration, we calculate
strength functions near the real axis. Though this is a more time-consuming
calculation, it allows us to locate beta-minus and beta-plus poles, determine if
there are errors in the contour integration, and decide how to correct incorrect
half-lives.

We identify two subsets of nuclei, shown in Fig.~\ref{fig:hl_corrections} panel
a), for which we perform this additional calculation. The first, indicated by
red circles, is a set of 224 odd nuclei that have decay rates significantly
below the average for a given $Q$-value or that contain significant negative
contributions. We refer to this set as ``suspicious.'' The second, shown with
blue squares, is a random sample of 100 odd nuclei from the remaining
population. Assuming that the probability of a half-life requiring correction is
uniformly distributed, this sample size allows us to estimate the proportion of
half-lives that require correction with a 10\% margin of error at a 95\%
confidence level. We find that more than half of the examined lifetimes turn out
to be correct, and those that are not contain errors of two types.

The first type, illustrated by the top panels of
Fig.~\ref{fig:contour_integration_failure}, can be corrected by simply shifting
the left bound of the contour. In the figure, the original left bound
(Eq.~\eqref{eq:contour_left_bound}) is the dashed vertical line, while the
corrected left bound is the solid vertical line. There are two situations which
cause this type of error. The first, similar to that shown in
Fig.~\ref{fig:contour_integration_failure} panel b), occurs when the HFB
estimate $\Omega_1$ is negative but the residual interaction moves it to
a positive number $E < \lvert \Omega_1 \rvert$. This is corrected by
placing the left bound at zero. The second, illustrated in panels a) and b) of
Fig.~\ref{fig:contour_integration_failure}, occurs when there exists a
beta-minus (beta-plus) transition at negative (positive) energy, but either the
corresponding beta-minus or beta-plus strength itself is negligible.  This
behavior occurs almost exclusively in odd nuclei adjacent to closed shells,
where pairing vanishes and the transition that takes the parent farther
from the closed shell is suppressed.  These cases are corrected by shifting the
contour to exclude (include) beta-minus (beta-plus) poles with negligible
strength.

The second type of error, exemplified by panel c) of
Fig.~\ref{fig:contour_integration_failure}, is more difficult to correct. Two
situations can give rise to this shape in the strength distribution: the
existence of a non-negligible beta-minus pole at negative energy and an
associated non-negligible beta-plus pole at positive energy, or, as in panels c)
and d), the existence of poles at imaginary energies. To determine if any
corrections are warranted, we pinpoint the location of the poles by calculating
the strength parallel to the imaginary axis out to 1~MeV.  We examine the
strength in each multipole, and if the original contour integration contains any
errors, we integrate along a contour that surrounds only the problematic
poles (and only them) to determine the correction.

We identify 60 nuclei --- 54 in the suspicious set and 6 in the random sample
--- that require only a simple adjustment of the contour, and 41 nuclei --- 26
in the suspicious set and 15 in the random sample --- that require more careful
corrections (33 of which have an imaginary pole in at least one multipole). The
results of correcting the half-lives appear in panel b) of
Fig.~\ref{fig:hl_corrections}. The amount of change is indicated by the black
arrowheads. Most of the arrowheads lie hidden beneath the circles or squares,
usually because the problems are in forbidden multipoles that contribute only a
small amount to the rate.  Only a few half-lives shrink by more than an order of
magnitude, when a low-lying beta-minus transition is missing from the original
contour. Some half-lives increase slightly after we remove positive
contributions from imaginary poles.  Orange triangles in panel b) correspond to
nuclei with negative total decay rates that became positive after correction. 

Our random sample suggests that about 6\% of our unexamined results should be
corrected simply, by shifting the left bound of the contour, and about 15\% may
require more intricate corrections.  Only a single half-life in the random
sample changes by more than 5\%, however (it changes by 30\%). Thus, the
corrections to unverified half-lives are very likely small compared to the
average error in our rates (see Fig.~\ref{fig:bayesian_result}).  Nuclei with
half-lives that require significant correction very probably belong to
the suspicious set that we have just analyzed.

Finally, we should mention that numerical error is an additional source of
small negative contributions to rates. Both the HFB and FAM solutions
contain numerical error from several sources, e.g., incomplete convergence,
truncation, etc. These errors are compounded in the final strength function and 
amplified by the phase space.  If a rate is very small, the contour
integral that generates it can suffer from incomplete cancellation of large 
oscillations.  In compiling our final table of half-lives, presented here as 
supplemental material~\cite{supplemental}, we break each rate into 
contributions from each multipole, set any negative contributions to zero, and 
re-sum. This procedure usually changes rates by less than $5\%$.

\begin{figure}[!bt]
\includegraphics[width=1\columnwidth]{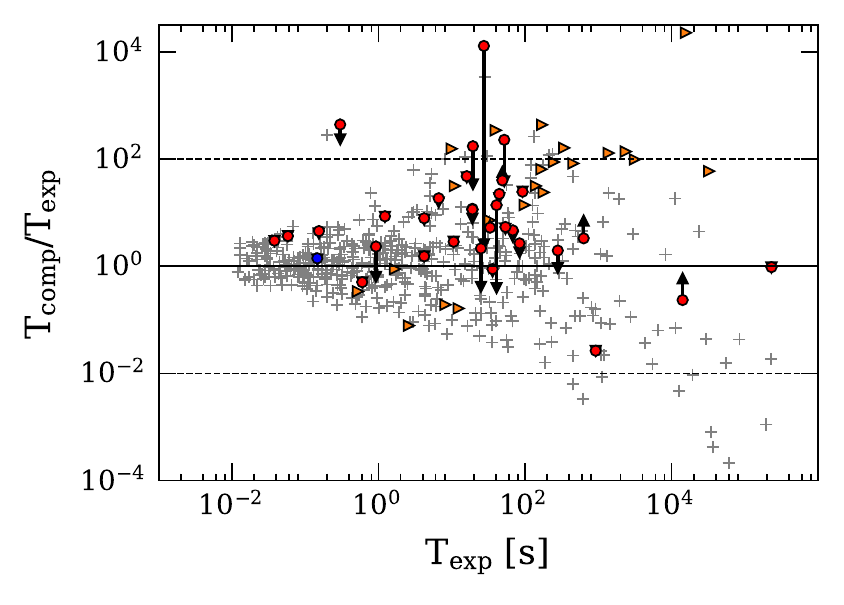}
\caption{\label{fig:ensdf_comparison_corrections} Same as panel b) of
Fig.~\ref{fig:hl_corrections} but compared with 2019 ENSDF data. Only odd nuclei
are shown.} \end{figure}

In Fig.~\ref{fig:ensdf_comparison_corrections} we compare our final results with
2019 ENSDF experimental data~\cite{ENSDF2019} for nuclei with experimental
half-lives less than $10^{6}$~s. We highlight half-lives that are corrected, as
in Fig.~\ref{fig:hl_corrections} panel b), and find that corrections almost
always improve the agreement with experiment. The majority of our data fall
within one or two orders of magnitude of experiment for half-lives less than
1000~s. In the next section, we will quantify more rigorously the theoretical
uncertainties associated with such calculations.

\begin{figure}[!hbt]
\includegraphics[width=1\columnwidth]{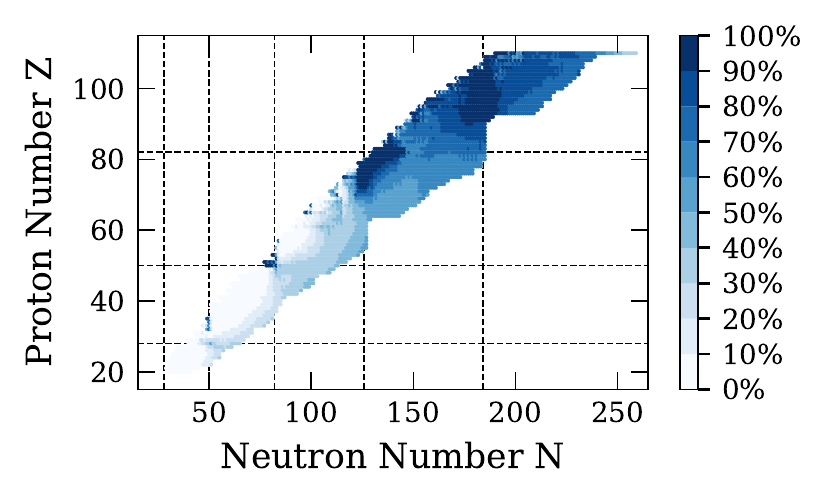}
\caption{\label{fig:first_forbidden_contributions} First-forbidden contribution
to the rates.  } \end{figure}

Figure~\ref{fig:first_forbidden_contributions} displays the contributions to
decay rates of first-forbidden operators. We find, as do other groups, that 
first-forbidden contributions are important in many nuclei and observe 
competing effects: forbidden contributions scale with the nuclear radius and 
$Q$-value, becoming important in heavier nuclei far from stability, but they 
also become important near stability and closed shells where the allowed rate 
is very small and allowed contributions are suppressed.

\subsection{\label{ssec:error}Error analysis}

One major challenge facing large scale calculations is the quantification of
uncertainty. Most of the nuclei considered here are not experimentally
accessible, and so we lack an experimental benchmark with which to evaluate our
calculations. A simple way to deal with this challenge is to develop a
\emph{model} for the error. The model can be fit to data where available, and
then extrapolated or interpolated to estimate errors for the remaining data. We
use the simple model developed in Ref.~\cite{Mustonen2016}, which we summarize
here.  The error parameter of interest is, for the $i^{th}$
nucleus~\cite{Moller2003}, 
\begin{equation}
\label{eq:error_parameter_r}
r_i = \log_{10} \bigg(\frac{t_{\text{th}}}{t_{\text{exp}}}\bigg) \,.  
\end{equation} 
To motivate a regression model for this parameter, we assume that there is a
single dominant transition to a state near the daughter ground state, and that
the forbidden shape factors depend much less on the $Q$-value than does the
allowed phase space. These assumptions allow us to assign a single effective
$Q$-value and shape factor $C_{\rm eff}$ to the decay; cf.\ \cite{Mustonen2014}
for the definition of the shape factor $C$. Using $q_{\rm eff}$ to denote the
effective $Q$-value in units of electron mass ($q_{\rm eff} = Q_{\rm eff} / m_e
c^2$), we model the error $r_i$ on the rate, as a function of the theoretical
$Q$-value and charge of the daughter nucleus, as, 
\begin{equation}
r_i(q_{\text{g.s.}}^{\text{th}},Z_f) \approx c_{r_i} +
f_r(q_{\text{g.s.}}^{\text{th}}+1,Z_f) q_{r_i} \,, 
\end{equation} 
where the errors in the effective shape factor, $c_{r}$, and the effective
$Q$-value, $q_{r}$, are defined by 
\begin{equation} 
c_r \equiv
\log_{10}\frac{C^{\text{exp}}_{\text{eff}}}{C^{\text{th}}_{\text{eff}}}, \quad
q_r \equiv \frac{q_{{\text{eff}}}^{{\text{exp}}} -
q_{\text{eff}}^{\text{th}}}{\ln10}\,,
\end{equation} 
and the $Q$-value dependence is carried by the phase space factor, 
\begin{equation} 
f_r(q+1,Z_f) \equiv \frac{1}{f(q+1,Z_f)} \frac{df(q+1,Z_f)}{dq} \,.
\end{equation}

Next, we assume that the $c_{r_i}$ and $q_{r_i}$, which depend on the nucleus
$i$, are each normally distributed random variables with widths that are
independent of the $Q$-value, and that the distributions for $c_{r_i}$ and
$q_{r_i}$ contain a systematic bias that is independent of the nucleus and the
$Q$-value.  These assumptions allow us to write the error parameters for nucleus
$i$ in the form 
\begin{equation} 
\begin{aligned} 
c_{r_i} &= b_c + \epsilon_{c}, \quad\epsilon_{c} \sim\mathcal{N}(0,\sigma_{c})\,,  \\ 
q_{r_i} &= b_q + \epsilon_{q}, \quad\epsilon_{c} \sim\mathcal{N}(0,\sigma_{q})\,, 
\end{aligned} 
\end{equation} 
where $b_{c}, b_{q}, \sigma_{c}, \sigma_{q}$ are still undetermined parameters.
Finally, since the assumptions of the model are best for large $Q$-values, we
can make use of the Primakoff-Rosen approximation to the allowed phase
space~\cite{Suhonen2007}, which lets us express $f_r(q+1,Z_f)$ as a simple
rational function with no explicit dependence on the charge $Z_f$ of the
daughter nucleus:
\begin{multline} 
f_r(q+1,Z_f) \approx f_{r}^P(q) \\ 
\equiv \frac{5 (q+1)^4 - 20 (q+1) + 15}{(q+1)^5 - 10 (q+1)^2 + 15 (q+1) - 6} \,.
\end{multline} 
We then end up with a one dimensional, non-linear error model with noise: 
\begin{equation}
r_i(q_{\text{g.s.}}^{\text{th}}) = b_c + f_{r}^{P}(q_{\text{g.s.}}^{\text{th}})b_{q}
+ \epsilon_i, \quad \epsilon_i \sim \mathcal{N}(0, \sigma_{r}) \,.
\end{equation}
Because $c_{r}$ and $q_r$ are independent, their widths add in quadrature.  We
find, however, that only $\sigma_q$ is important and therefore take the width of
the total noise term to be
\begin{equation} 
\sigma_r(q_{\text{g.s.}}^{\text{th}}) = \sqrt{\sigma_c^2 +
(f_r^P(q_{\text{g.s.}}^{\text{th}}) \sigma_q)^2} \approx
f_r^P(q_{\text{g.s.}}^{\text{th}}) \sigma_q \,.  
\end{equation} 
That leaves three unknown parameters $b_c, b_q, \sigma_q$ to be determined.

\begin{figure}
\includegraphics[width=1\columnwidth]{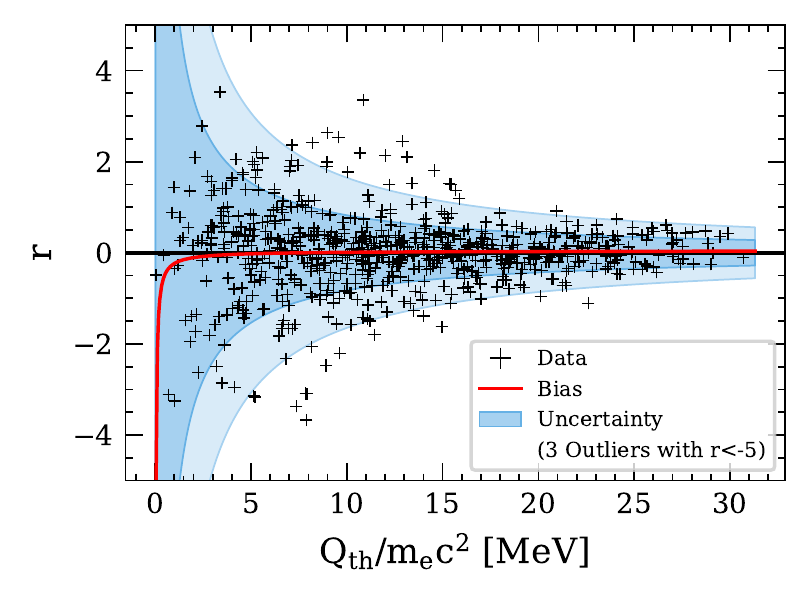}
\caption{\label{fig:bayesian_result} Bayesian fit to the bias function and
one- and two-standard-deviation bands.} 
\end{figure}

To estimate the parameters, we use our own python adaptation of a Metropolis
Monte Carlo code from Ref.~\cite{bailer-jones2017} to sample the unnormalized
Bayesian posterior distributions of ${\beta = \tan^{-1}{b}}$ and $\sigma_q$,
with priors
\begin{equation}
\label{eq:priors} 
\begin{aligned}
P(\beta_{c}) &= P(\beta_{q}) = \frac{1}{2\pi} \\ 
P(\sigma_{q}) &\propto \log(\sigma_{q}) \,.
\end{aligned}
\end{equation} 
The sampling probability distribution is a multivariate Gaussian with a variance
of $(0.02)^2$ for all three parameters. Following a burn-in period of $200,000$
steps, we retain every 100$^{\text{th}}$ iteration from the next million steps
to reduce autocorrelation. From Gaussian kernel density estimates of the
resulting distributions we esimate the most likely values to be $b_c=0.049$,
$b_q=-0.082$, and $\sigma_q=1.807$.  Figure~\ref{fig:bayesian_result} shows the
resulting confidence regions on top of our entire data set. We find hardly any
bias, indicating that our half-lives are equally likely to be over- and
under-predicted. The model is not reliable for very small $Q$-values, but for
moderate to large $Q$-values it predicts that the majority of our calculated
half-lives will differ from experiment by less than one order of magnitude. The
data is slightly non-Gaussian, with the one and two standard deviation bands
capturing $76\%$ and $94\%$ of the 718 data points, respectively.

\subsection{\label{ssec:comparisons}Comparisons}

\begin{figure}[!htb]
\includegraphics[width=\columnwidth]{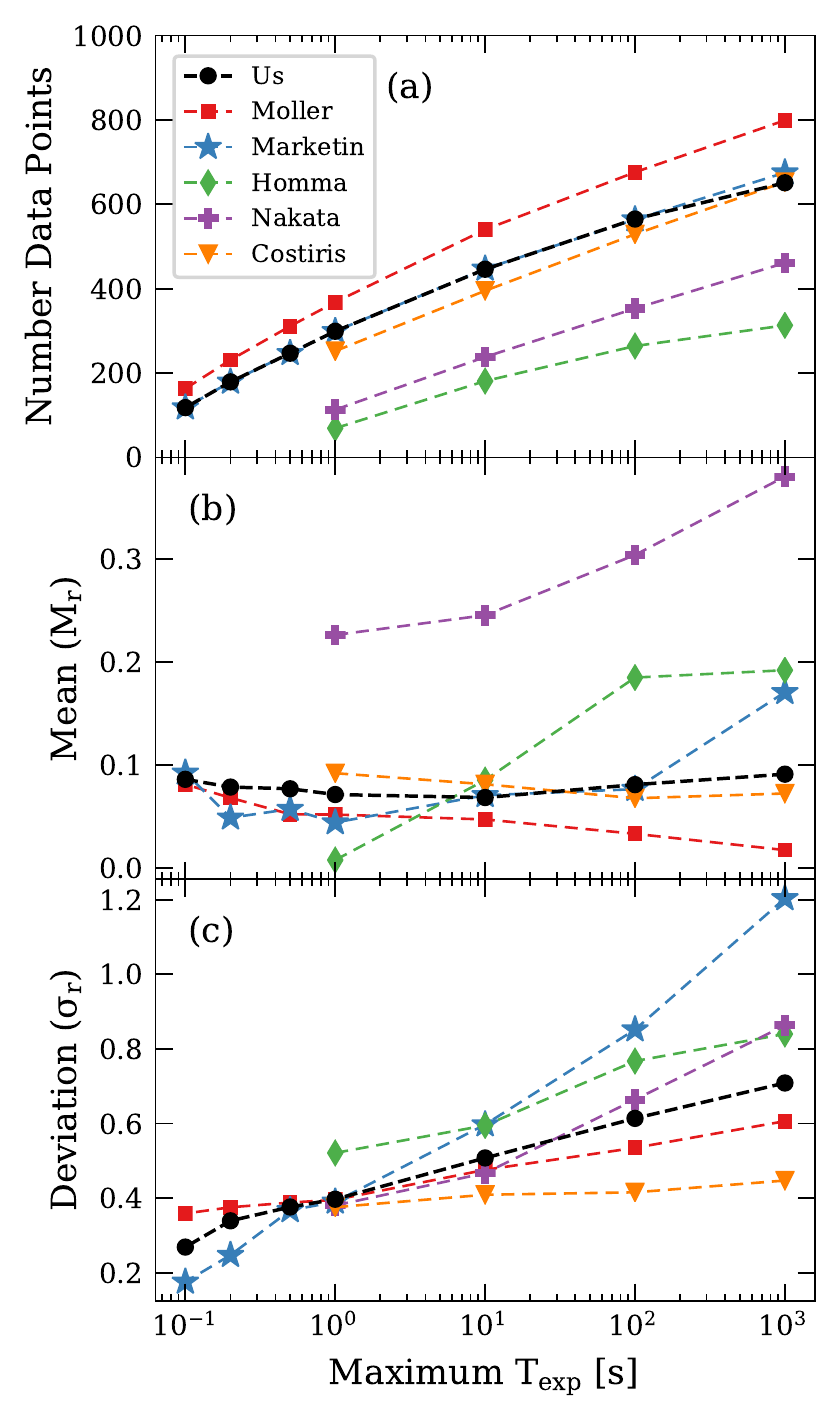}
\caption{\label{fig:quality_measures} Comparison of {}error-evaluation parameters
among results of Refs.~\cite{Marketin2016} (Marketin),~\cite{Moller2003}
(M\"oller),~\cite{Costiris2009} (Costiris),~\cite{Nakata1997} (Nakata), and
\cite{Homma1996} (Homma).}
\end{figure}

To evaluate our data where experimental values are unavailable, we compare our
results to those of other global beta-decay calculations. The authors of
Ref.~\cite{Homma1996} (labeled ``Homma'' in Fig.~\ref{fig:quality_measures})
conducted a microscopic pnQRPA calculation with schematic allowed and unique
first-forbidden interactions, and treated odd nuclei perturbatively.
Reference~\cite{Nakata1997} (labeled ``Nakata'') carried out a macroscopic
calculation within the semi-gross theory.  Reference~\cite{Moller2003} (labeled
M\"oller) combined microscopic and macroscopic approaches, using the
finite-range droplet model for ground state properties, the pnQRPA with an
empirical spreading for Gamow-Teller strength, and the gross theory for
first-forbidden contributions.  More recently, Ref.~\cite{Costiris2009} (labeled
``Costiris'') applied a neural network to predict half-lives.  Finally,
Ref.~\cite{Marketin2016} (labeled ``Marketin'') conducted a fully
self-consistent covariant pnQRPA calculation with local fits to the isoscalar
pairing strength, treating odd nuclei as if they were fully paired even nuclei
with an odd number of nucleons on average. 

\begin{figure*}[t]
\includegraphics[width=2\columnwidth]{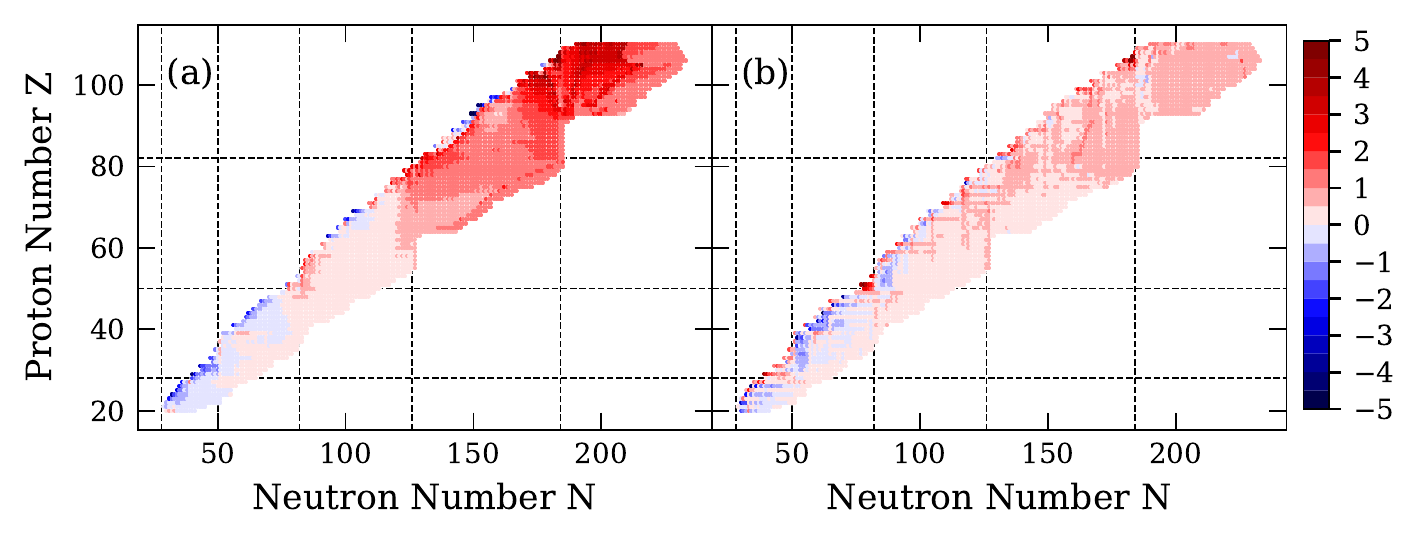} 
\caption{\label{fig:nuclear_chart_hl_comparisons} Log of the ratio of our
half-lives to those of a) Ref.~\cite{Marketin2016} and b)
Ref.~\cite{Moller2003}.  } 
\end{figure*}

To compare our results to those of the other papers, we use the quality measures
outlined, e.g., in Ref.~\cite{Moller2003}: the mean $(M_r)$ and standard
deviation $(\sigma_r)$ of the error parameter in Eq.~\eqref{eq:error_parameter_r},
\begin{equation} \label{eq:quality_measures} 
M_r = \frac{1}{n} \sum\limits_{i=1}^n r_i, \quad \sigma_r = \bigg[ \frac{1}{n}
\sum\limits_{i=1}^n (r_i - M_r)^2 \bigg]^{1/2}\,.
\end{equation}
We present these measures for the set of nuclei with experimental half-lives
less than 1000~s, 100~s, 1~s, 0.5~s, 0.2~s, and 0.1~s.  For
Refs.~\cite{Homma1996,Nakata1997,Costiris2009} we take the measures directly
from the corresponding paper.  References~\cite{Moller2003,Marketin2016}
supplied their data set as supplemental material, and we recompute the quality
measures with the more recent 2019 ENSDF experimental
half-lives~\cite{ENSDF2019}. Figure~\ref{fig:quality_measures} summarizes the
results.  The differences in experimental data sets considered in each paper can
be seen in part by noting the number of data points used to compute the quality
measures. The errors for Ref.~\cite{Marketin2016} are somewhat larger for
long-lived isotopes than the values given in that paper because we include all
the calculations in odd nuclei, while the authors excluded a few that they
considered outliers.  In general, our calculation is comparable in fidelity to
the others.  Unlike those, however, its treatment of odd nuclei is fully
self-consistent, capturing in part the one-quasiparticle nature of such states
through the EFA, and it uses a single energy functional with no local
adjustments.  Figure~\ref{fig:nuclear_chart_hl_comparisons} compares all our
results with those provided in Refs.~\cite{Moller2003,Marketin2016}.  We
generally predict longer half-lives than the other two models in heavier nuclei,
and slightly shorter half-lives in lighter nuclei. The vast majority of our
numbers fall within one order of magnitude of those of Ref.~\cite{Moller2003}.
Both we and Ref.~\cite{Moller2003} predict significantly longer half-lives in
heavy isotopes than does Ref.~\cite{Marketin2016}. There do not appear to be any
other significant systematic differences among the results.

\section{\label{sec:conclusions}Conclusions}

Using the statistical extension of the charge-changing finite amplitude method,
we computed beta-decay half-lives of almost all odd-mass and odd-odd nuclei on
the neutron-rich side of stability, in a fully microscopic and self-consistent
way. The equal filling approximation allows us to retain time-reversal symmetry
while sill largely including the effects of core polarization by the odd
nucleon. We showed that in a few cases the EFA leads to the appearance of
negative and even imaginary eigenvalues. Overall our half-lives are similar to
those of other global calculations in reproducing experimental data.  We
supplemented these calculations with an estimate of theoretical uncertainties,
which suggest that calculated half-lives fall within two orders of magnitude of
experimental values for nuclei with $Q$-values greater than about 2~MeV. We also
find, as do other groups, that first-forbidden contributions are important in
many nuclei. We provided all the half-lives described here, along with
associated ground-state properties, error estimates, and Gamow-Teller strength
distributions, in the supplemental material~\cite{supplemental}.

We plan to extend our methods in several ways: 
\begin{itemize}
\item We will use the statistical FAM with the grand canonical ensemble for
finite temperature beta-decay calculations. Decay at non-zero temperature plays
an important role in neutron-star mergers and core-collapse
supernovae~\cite{Langanke2000,Langanke2001}. 
\item We will improve the ability of the FAM to capture low-energy strength by
including correlations beyond the QRPA. Although one must be careful in
combining such correlations with density functionals, several procedures exist
for doing so~\cite{Gam15,Robin16,Niu2018}. An efficient implementation of an
extension to the FAM would allow better global calculations.  
\item Finally, we will better treat the weak interaction.  Here we restrict
ourselves to the impulse approximation, neglecting many-body currents
completely.  Recent work shows that such currents account for a significant
fraction of the quenching of Gamow-Teller strength~\cite{Gysbers2019}.  With an
additional extension of the pnFAM we can take two-body currents into account.
\end{itemize}
Our calculations are also an important milestone in the development of a
consistent description of the fission process within nuclear
DFT~\cite{schunck2016}.  Although spontaneous fission-fragment half-lives,
fragment distributions, and fragment excitation energies can already be computed
in DFT, our work paves the way to for a description of the deexcitation of the
fragments, including gamma emission and beta decay, within the same framework.

\section*{\label{sec:acknowledgments}Acknowledgments}
Many thanks to M. Mustonen and T. Shafer, for guidance on the pnFAM, and to S.
Guilliani for helpful discussions on beta decay. This work was supported in part
by the Nuclear Computational Low Energy Initiative (NUCLEI) SciDAC-4 project
under U.S.\ Department of Energy grant DE-SC0018223 and the FIRE collaboration.
Some of the work was performed under the auspices of the U.S.\ Department of 
Energy by Lawrence Livermore National Laboratory under Contract 
DE-AC52-07NA27344. Computing support came from the Lawrence Livermore National 
Laboratory (LLNL) Institutional Computing Grand Challenge program.


%
\end{document}